\documentclass{article}
\usepackage{amsmath}
\usepackage{amsfonts}
\usepackage{amsthm}
\usepackage{amssymb}
\usepackage{url}
\usepackage{hyperref}

\input{Qcircuit}

\newcommand{\proj}[2]{\ket{#1}\!\bra{#2}}

\newcommand{\braket}[2]{\left \langle #1 | #2 \right \rangle}
\newcommand{\tr}[2][ ]{\text{Tr}_{#1}\!\left( #2 \right)}
\newcommand{\norm}[1]{\left|\left|#1\right|\right|}

\newtheorem{definition}{Definition}
\newtheorem{theorem}{Theorem}
\newtheorem{lemma}{Lemma}

\title{On the power quantum computation over real Hilbert spaces}
\author{Matthew McKague \\ Department of Computer Science \\ University of Otago \\
\url{matthew.mckague@otago.ac.nz}}

\begin{document}
\maketitle
\begin{abstract}
We consider the power of various quantum complexity classes with the restriction that states and operators are defined over a real, rather than complex, Hilbert space.  It is well know that a quantum circuit over the complex numbers can be transformed into a quantum circuit over the real numbers with the addition of a single qubit.  This implies that $\mathsf{BQP}$ retains its power when restricted to using states and operations over the reals.  We show that the same is true for $\mathsf{QMA}(k)$, $\mathsf{QIP}(k)$,  $\mathsf{QMIP}$, and $\mathsf{QSZK}$.

\end{abstract}

%%%%%%%%%%%%%%%%%%%%%%%%%%%%%%%%%%%%%%%%%%%%%%
\section{Introduction}

The standard quantum formalism specifies a complex Hilbert space, but one could just as well consider a real Hilbert space.  In fact, a quantum formalism based on real Hilbert spaces has the same descriptive power as complex quantum formalism:  given any description of a physical system using complex quantum formalism, there is a simple mapping to a description over real Hilbert spaces, called the \emph{real simulation}, that gives the same predicted outcome statistics.  Moreover, this mapping respects the division into subsystems  \cite{McKague:2009:Simulating-Quan}.  This means, for example, that there can be no experiment that can rule out real quantum formalism.\footnote{Unless one makes assumptions about the dimension of systems.}

Now let us consider computational complexity.  Does the choice of real or complex Hilbert spaces change the power of various computing models?  At first it might seem that this is obvious:  if any quantum system can be described using real Hilbert spaces, then why not the physical systems underlying some computing model?  For computing models where all parties are trusted this is indeed the case.  $\mathsf{BQP}$, $\mathsf{EQP}$ and related classes are all unchanged as well as classes where messages are classical, such at $\mathsf{QCMA}$.

In the case of interactive proofs, however, the situation is more complicated.  Mapping some complex interactive proof protocol into a real proof protocol gives the same completeness, but it might give more power to a cheating prover.  As well, even if we lift a real proof protocol into a complex protocol there are more operations available to a cheating prover.  Hence, on the face of it, there is no obvious relationship between complexity classes defined over complex or real Hilbert spaces.  However, we will show that, in many interesting cases, restricting to real Hilbert spaces does not change the power of a complexity class.

%%%%%%%%%%%%%%%%%%%%%%%%%
\subsection{Definitions}

For an introduction to complexity theory and interactive proofs, see \cite{Arora:2009:Computational-C}.  Here we recall the definitions of several complexity classes.
\begin{definition}
$ \mathsf{QMIP}[m,k]$\cite{Kobayashi:2001:Quantum-Multi-P} is the set of languages which can be verified by a polynomial time quantum verifier interacting with $m$ non-communicating quantum provers, exchanging $k$ messages, which accepts inputs in the language with probability at least $2/3$ (completeness) and accepts inputs not in the language with probability no more than $1/3$ (soundness).  Further,
\begin{itemize}
	\item $\mathsf{QMIP}_{ne}[m,k]$ is the same as $\mathsf{QMIP}[m,k]$ with the additional restriction that the provers do not share prior entanglement.
	\item $\mathsf{QIP}(k) = \mathsf{QMIP}[1,k]$ \cite{Watrous:1999:PSPACE-has-2-ro}
	\item $\mathsf{QMA} = \mathsf{QIP(1)}$ \cite{Watrous:2000:Succinct-quantu}
	\item $\mathsf{QMA}(k) = \mathsf{QMIP}_{ne}[k,1]$ \cite{Kobayashi:2001:Quantum-Certifi}
\end{itemize}
%¥
\end{definition}
%¥

Note that the completeness and soundness values can be set to any constants, with completeness larger than soundness, without changing the complexity classes.  Also, since there is no advantage in the verifier sending the final message, by convention the provers always send the final message, so that if $k$ is odd the provers send the first message, otherwise the verifier sends the first message.

\begin{definition}
$\mathsf{QSZK}$ \cite{Watrous:2002:Limits-on-the-p} is the subset of languages $\mathsf{QIP}$ with the additional restriction that, for an honest prover, the state $\rho_{x,j}$ consisting of the verifier's state and all message registers after $j$ rounds of interaction of the proof system with input $x$ satisfies
\begin{equation}
\norm{\rho_{x,j} - \sigma_{x,j}} \leq \delta(|x|)
\end{equation}
for all $x$ in the language, and all $j$ or some function $\delta$ which is bounded above by $1/p(n)$ for all polynomials $p(n)$ for sufficiently large $n$, where $\sigma_{x,j}$ are the outputs of a uniform family of polynomially sized circuits.  
\end{definition}
%¥

Essentially, this definition says that a verifier does not gain any information by interacting with the prover that it could not have discovered through a polynomial-time computation on its own.  Note that $\mathsf{QSZK}$ does not change if we restrict to honest verifiers or we allow them to deviate from the protocol\cite{Watrous:2009:Zero-Knowledge-}. 

For any quantum complexity class let us define a restricted class such that all states and operators are defined over the real numbers instead of the complex numbers.  We name these classes by adding a subscript $\mathbb{R}$.  Thus $\mathsf{BQP}$ restricted to real gates and states becomes $\mathsf{BQP}_{\mathbb{R}}$, etc..  We also place the same restriction on provers for the interactive complexity classes.

We might also consider the case where the provers are allowed to be complex.  However, it is our intention to consider the computational complexity of various classes where the entire underlying vector space is real, rather than test the power of a real verifier.  However, as a consequence of our work it does not change the power if the prover is allowed to be real or complex.  Indeed, when considering cheating provers we will frequently relax the definition to allow complex provers as a simplification.

%%%%%%%%%%%%%%%%%%%%%%%%%
\subsection{Contributions}
 We prove the following theorem:
\begin{theorem}
\begin{eqnarray}
\mathsf{QMA}  &= &\mathsf{QMA}_{\mathbb{R}} \\
\mathsf{QMA}(k) & = &\mathsf{QMA}_{\mathbb{R}}(k)\\ 
\mathsf{QIP}(k) & = &  \mathsf{QIP}_{\mathbb{R}}(k)\\ 
\mathsf{QMIP}[m, k] & = &  \mathsf{QMIP}_{\mathbb{R}}[m, k]\\  
\mathsf{QMIP_{ne}}[m, k] & \subseteq &  \mathsf{QMIP_{ne}}_{\mathbb{R}}[m, k+1]\\ 
\mathsf{QMIP_{ne}}_{\mathbb{R}}[m, k] & \subseteq &  \mathsf{QMIP_{ne}}[m, k+1]\\ 
\mathsf{QSZK} & = &  \mathsf{QSZK}_{\mathbb{R}}.  
\end{eqnarray}
\end{theorem}

For each class $C$ we must prove both containments: $C \subseteq C_{\mathbb{R}}$ and $C_{\mathbb{R}} \subseteq C$.  For the first case we use the real simulation, which preserves completeness.  We show that soundness is also preserved by using an argument from \cite{McKague:2011:Generalized-Sel}, originally used for security.  To show $C_{\mathbb{R}} \subseteq C$ we lift a real protocol into analogous complex Hilbert spaces.  This preserves completeness directly.  To show that soundness is also preserved we apply the real simulation to a cheating prover.

In addition to the basic arguments above, there are several special cases.  For $\mathsf{QMA(2)}$, we cannot use the usual real simulation because it introduces entanglement between provers.  For this we must develop a different real simulation which does not require entanglement.  A similar problem, but different solution, occurs for $\mathsf{QMIP_{ne}}$.  Finally, for $\mathsf{QSZK}$ we must take into account the additional requirement that an honest prover does not leak too much information to a verifier.  This requires us to reverse the real simulation process and construct an efficient complex circuit from a real circuit. 

%%%%%%%%%%%%%%%%%%%%%%%%%%%%%%%%%%%%%%%%%%%%%%
\section{Real simulation}
In this section we recall the relevant work on real simulation of arbitrary quantum systems over complex Hilbert spaces.  In particular, we recall the real simulation of quantum circuits and multi-party computations.  Additionally, we introduce the separable simulation for the case where the original multi-party measurement is separable.  Finally, we introduce the ``reverse'' mapping to obtain complex circuits from real circuits.

%%%%%%%%%%%%%%%%%%%%%%%%%
\subsection{Single party}\label{sec:singlesystem}
It is a well known result that any quantum circuit can be simulated over a real Hilbert space by using one additional qubit.  The basic idea is to store the 2-dimensional real vector space represented by a complex number in a 2-dimensional real vector space corresponding to the additional qubit added to the circuit.  It is quite easy to derive the transformation needed to take states and gates over a complex Hilbert spaces to states and gates over real Hilbert space.

\begin{definition}
Let $\mathcal{A}$ be a complex Hilbert space, and $\mathcal{Q} = \mathcal{H}_{2}$ be a 2-dimensional Hilbert space.  Let $R: \mathcal{A} \rightarrow \mathcal{Q} \otimes \mathcal{A}$ be defined by
\begin{equation}
R(\ket{\psi}_{\mathcal{A}}) = \ket{0}_{\mathcal{Q}} \text{Re} \ket{\psi}_{\mathcal{A}} + \ket{1}_{\mathcal{Q}} \text{Im} \ket{\psi}_{\mathcal{A}}.
\end{equation}
where $\mathcal{Q} \otimes \mathcal{A}$ is taken as a real Hilbert space.  Further, $R$ is extended to the dual space by $R(\bra{\psi}) = R(\ket{\psi})^{\dagger}$ and to matrices by $R(\proj{\psi}{\phi}) = R(\ket{\psi})R(\bra{\phi})$ and by the fact $R(aM + bN) = aR(M) + bR(N)$ where $a,b \in \mathbb{R}$ and $M,N$ are complex matrices.
\end{definition}
%¥

$R(\cdot)$ has several interesting properties, summarized in the following lemma.
\begin{lemma}\label{lemma:propertiesofR}
Let $M$, $N$ be linear operators over a Hilbert space $\mathcal{A}$ and $\ket{\psi}, \ket{\phi} \in \mathcal{H}$. Further, define $V_{\mathcal{Q}}$ by
\begin{equation}
V_{\mathcal{Q}} = \frac{1}{\sqrt{2}} \left(
	\begin{matrix}
	1 & 1 \\
	-i & i \\
	\end{matrix}
\right)_{\mathcal{Q}}.
\end{equation}
Then
\begin{eqnarray}
\label{eq:vbasis}
V^{\dagger}_{\mathcal{Q}} R(M) V_{\mathcal{Q}} & = & \proj{0}{0}_{\mathcal{Q}} \otimes M_{\mathcal{A}} + \proj{1}{1}_{\mathcal{Q}} \otimes M^{*}_{\mathcal{A}} \\
V^{\dagger}_{\mathcal{Q}} R(\ket{\psi}) & = & \frac{1}{\sqrt{2}}\ket{0}_{\mathcal{Q}} \ket{\psi}_{\mathcal{A}} + \frac{1}{\sqrt{2}}\ket{1}_{\mathcal{Q}} \ket{\psi^{*}}_{\mathcal{A}} \\
R(MN) & = & R(M) R(N)Ê\\
R(M \ket{\psi}) & = & R(M) R(\ket{\psi}) \\
R(\bra{\psi}) R(M) R(\ket{\phi}) & = & \text{Re} \bra{\psi} M \ket{\phi}
\end{eqnarray}
%¥

Also,  $R(M)$ is unitary (Hermitian, positive semi-definite) if and only if $M$ is unitary (Hermitian, positive semi-definite).

\end{lemma}
%¥

For further details, including proofs, see \cite{McKague:2010:Quantum-Informa}.  A straightforward consequence of the lemma is that if we transform the input state, each unitary and each measurement in a circuit by $R(\cdot)$, then we obtain the same distribution on measurement outcomes as in the original circuit.  The circuit thus obtained is called the \emph{real simulation}.  Using the real simulation we learn that $\mathsf{BQP}_{\mathbb{R}} = \mathsf{BQP}$.  The cost incurred for using the simulation is one qubit.  Note, however, that this ``extra'' qubit $\mathcal{Q}$ has a rather important role, and is needed to perform any unitary $R(U)$ when $U$ has complex entries.

Although it is not necessary for the simulation to work, we introduce the unitary $V_{\mathcal{Q}}$ since it will be useful later on.  By changing the basis by $V_{\mathcal{Q}}$ we can view the real simulation as a coherent mixture of the original computation and its complex conjugate.  Also, in this basis it is quite obvious that the real simulation produces the same statistics as the original circuit.

%%%%%%%%%%%%%%%%%%%%%%%%%
\subsection{Multiple parties}\label{sec:multipartite}
Suppose that we have some computation that happens between multiple parties and we wish to simulate the computation over real numbers instead of complex.  Furthermore, we need the simulation to respect the original division into multiple parties.  A good example is Bell tests, where we wish to show that real states and measurements can violate a Bell inequality by the same amount as complex states and measurements.  To accomplish this, we give a ``copy'' of $\mathcal{Q}$ to each party.

\begin{definition}
For $j = 1 \dots m$ let $\mathcal{A}_{j}$ be a complex Hilbert spaces, and $\mathcal{Q}_{j} = \mathcal{H}_{2}$ be a 2-dimensional Hilbert spaces.  Let $R^{(m)}: \bigotimes_{j} \mathcal{A}_{j} \rightarrow \bigotimes_{j} \mathcal{Q}_{j} \otimes \mathcal{H}_{j}$ be defined by
\begin{multline}
R^{(m)}(\ket{\psi})  =  V_{\mathcal{Q}_{1}} \otimes \dots \otimes V_{\mathcal{Q}_{m}} \left(
 \frac{1}{\sqrt{2}}\ket{00 \dots 0}_{\mathcal{Q}_{1} \dots \mathcal{Q}_{m}} 
\ket{\psi}_{\mathcal{A}_{1} \dots \mathcal{A}_{m}}  + \right.
\\
\left.  \frac{1}{\sqrt{2}}
\ket{11 \dots 1}_{\mathcal{Q}_{1} \dots \mathcal{Q}_{m}} 
\ket{\psi^{*}}_{\mathcal{A}_{1} \dots \mathcal{A}_{m}}
\right)
\end{multline}
and analogously for $R^{(m)}(\bra{\psi})$.  For operators we define $R^{(m)}_{j}(\cdot)$ by
\begin{equation}
V^{\dagger}_{\mathcal{Q}_{j}} R^{(m)}_{j} (M_{\mathcal{A}_{j}})  V_{\mathcal{Q}_{j}} = \proj{0}{0}_{\mathcal{Q}_{j}} \otimes M_{\mathcal{A}_{j}} + \proj{1}{1}_{\mathcal{Q}_{j}}\otimes M^{*}_{\mathcal{A}_{j}}
\end{equation}
\end{definition}
%¥

From the definition it is immediately clear that transforming a circuit by $R^{(m)}$ preserves statistics, since it essentially produces a mixture of the original circuit and its complex conjugate.  As well, $R^{(m)}$ and $R^{(m)}_{j}$ produce states and operators with real amplitudes and matrix entries.  We obtain the following properties of $R^{(m)}(\cdot)$.
\begin{lemma}\label{lemma:propertiesofRm}
Let $j,k \in \{1, \dots, m\}$ $M_{\mathcal{A}_{j}}$, $N_{\mathcal{A}_{k}}$ be linear operators.  Further, let $\ket{\psi} \in \bigotimes_{j}\mathcal{A}_{j}$.  Then 
\begin{eqnarray}
R^{(m)}(M_{\mathcal{A}_{j}} N_{\mathcal{A}_{k}}\ket{\psi}) 
& = & 
R^{(m)}_{j}(M_{\mathcal{A}_{j}}) R^{(m)}_{k}(N_{\mathcal{A}_{k}})R^{(m)}(\ket{\psi}) 
\\
R^{(m)}(\bra{\psi}) R^{(m)}_{j}(M_{\mathcal{A}_{j}}) R^{(m)}(\ket{\psi}) 
& = & 
\text{Re} \bra{\psi} M_{\mathcal{A}_{j}} \ket{\psi}
\end{eqnarray}
%¥

$R_{j}^{(m)}(M_{\mathcal{A}_{j}})$ is unitary (Hermitian, positive semi-definite) if and only if $M_{\mathcal{A}_{j}}$ is unitary (Hermitian, positive semi-definite).  Finally, $R^{(m)}(\ket{\psi})$ is a real vector, and $R^{(m)}_{j}(M_{\mathcal{A}_{j}})$ is a matrix with real entries.
\end{lemma}
%¥

For detailed proofs of these facts, we refer the reader to \cite{McKague:2010:Quantum-Informa} section 2.5.4.  The proof proceeds in a fashion similar to that of lemma~\ref{lemma:propertiesofR}.

%%%%%%%%%%%%%%%%%%%%%%%%%
\subsection{Separable states and operations}\label{sec:separable}
For this section we focus on states and measurement observables only, which is sufficient for our application in section~\ref{sec:qma}.  The techniques should be adaptable to separable superoperators as well.

Suppose that we have a computation with multiple subsystems which is separable.  That is, the state and observables can all be written in the form
\begin{equation}
M = \sum_{k} M_{\mathcal{A}_{1},k} \otimes \dots \otimes M_{\mathcal{A}_{m},k}
\end{equation}
where each $M_{\mathcal{A}_{j},k}$ and $M$ are all Hermitian.  For pure states, we require that the corresponding density matrix is separable, i.e. that the state is a product state.  To simulate general multi-partite computations the construction given in section~\ref{sec:multipartite} is required, and in particular this means that the state will be entangled across subsystems.  However, in the case where the state and measurement are separable, the entanglement is not necessary.

We will use similar methodology to that in the previous section, by considering coherent mixtures of the original and complex conjugate computation.  The structure of the separable operator allows us to complex conjugate only one subsystem in a well defined way.  In the case of many subsystems, we can conjugate on a subset which we specify by way of a bit-string.
\begin{definition}[Partial complex conjugation]
Let $M = \sum_{k} M_{\mathcal{A}_{1},k} \otimes \dots \otimes M_{\mathcal{A}_{m},k}$ be a separable operator.  Then
\begin{equation}
M^{*_{j}} = \sum_{k} M_{\mathcal{A}_{1},k} \otimes \dots \otimes M_{\mathcal{A}_{j},k}^{*} \otimes \dots \otimes M_{\mathcal{A}_{m},k}.
\end{equation}
Let $\ket{\psi} = \ket{\psi_{1}} \dots \ket{\psi_{m}}$ be a product state.  Then
\begin{equation}
\ket{\psi}^{*_{j}} = \ket{\psi_{1}} \dots \ket{\psi_{j}}^{*}Ê\dots\ket{\psi_{m}} .
\end{equation}
Further, define $M^{*_z}$ for $z \in \{0,1\}^{m}$ to be $M$ with $(\cdot)^{*_{k}}$ applied for each $j$ such that $z_{j} = 1$  (Note that the order does not matter) and analogously for $\ket{\psi}^{*_{z}}$.

\end{definition}

Just as the complex conjugation preserves outcome statistics, partial complex conjugation also preserves statistics.

\begin{lemma}\label{lemma:separableconjugation}
Let $M = \sum_{j} M_{\mathcal{A}_{1},j} \otimes \dots \otimes M_{\mathcal{A}_{m},j}$ and $N = \sum_{k} N_{\mathcal{A}_{1},k}\otimes \dots \otimes N_{\mathcal{A}_{m},k}$ be separable operators.  Then
\begin{equation}
\tr{MN} = \tr{M^{*_{z}} N^{*_{z}}}
\end{equation}
for all $z \in \{0,1\}^{m}$.
\end{lemma}
\begin{proof}
Note that it suffices to show $\tr{MN} = \tr{M^{*_{t}} N^{*_{t}}}$ for $t \in \{1, \dots, m\}$ since we may apply induction to obtain the full result.  As well, it suffices to show the result for $m = 2$ and $t = 1$ since we may permute systems and combine $m-1$ of the systems into one to reduce to this case.

\begin{eqnarray}
\tr{MN}  & = & \tr{\sum_{j,k} M_{\mathcal{A}_{1},j} \otimes M_{\mathcal{A}_{2},j} N_{\mathcal{A}_{1},k} \otimes N_{\mathcal{A}_{2},k}} \\
& = & \sum_{j,k} \tr{M_{\mathcal{A}_{1},j} N_{\mathcal{A}_{1},k}} \tr{M_{\mathcal{A}_{2},j} N_{\mathcal{A}_{2},k}} \\
& = & \sum_{j,k} \tr{M_{\mathcal{A}_{1},j} N_{\mathcal{A}_{1},k}}^{*} \tr{M_{\mathcal{A}_{2},j} N_{\mathcal{A}_{2},k}} \\
& = & \sum_{j,k} \tr{(M_{\mathcal{A}_{1},j})^{*} (N_{\mathcal{A}_{1},k})^{*}}\tr{M_{\mathcal{A}_{2},j} N_{\mathcal{A}_{2},k}} \\
& = & \tr{\sum_{j,k} (M{\mathcal{A}_{1},j})^{*} \otimes M_{\mathcal{A}_{2},j} (N_{\mathcal{A}_{1},k})^{*} \otimes N_{\mathcal{A}_{2},k}} \\
& = & \tr{M^{*_{\mathcal{A}_{1}}} N^{*_{\mathcal{A}_{1}}}}.
\end{eqnarray}
Here the third line follows from the fact that $M_{\mathcal{A}_{1},j}$ and $N_{\mathcal{A}_{1},k}$ are Hermitian, hence complex conjugation does not change the trace.

\end{proof}

We are now ready to define the real simulation for a separable computation.  We will consider the case of two subsystems, but the same techniques generalize for any number of subsystems.

\begin{definition}
Define 
$R_{1,2}^{(2)}: \mathcal{A}_{1} \otimes \mathcal{A}_{2} 
\rightarrow 
\mathcal{Q}_{1} \otimes \mathcal{A}_{1} \otimes \mathcal{Q}_{2} \otimes \mathcal{A}_{2}
$ by
\begin{equation}
R_{1,2}^{(2)}(\ket{\psi_{1}}\ket{\psi_{2}}) = R^{(2)}_{1}(\ket{\psi_{1}}) \otimes R^{(2)}_{2}(\ket{\psi_{2}})
\end{equation}
Where $R^{(2)}_{1}$ and $R^{(2)}_{2}$ are both $R$ as in section~\ref{sec:singlesystem}, with $R^{(2)}_{j}: \mathcal{A}_{j} \rightarrow \mathcal{Q}_{j} \otimes \mathcal{A}_{j}$.  For operators we define
\begin{equation}
R_{1,2}^{(2)}(M)  =  \sum_{k} R^{(2)}_{1}(M_{\mathcal{A}_{1},k}) \otimes R^{(2)}_{2}(M_{\mathcal{A}_{2},k}) \end{equation}

\end{definition}
%¥

We obtain the following properties:
\begin{lemma}\label{lemma:propertiesofRmsep}
Let $M$ be a separable operator over a Hilbert space $\mathcal{A} = \mathcal{A}_{1}Ê\otimes \mathcal{A}_{2}$.  Further, let $\ket{\psi} = \ket{\psi_{1}}\ket{\psi_{2}} \in \mathcal{A}$.  Then
\begin{eqnarray}
\sum_{z \in \{0,1\}^{2}} \proj{z}{z}_{\mathcal{Q}_{1}\mathcal{Q}_{2}} \otimes M^{*_{z}}_{\mathcal{A}}
& = & 
\left(V_{\mathcal{Q}_{1}}^{\dagger}\otimes V_{\mathcal{Q}_{2}}^{\dagger}\right)R^{(m)}_{1,2}(M) \left(V_{\mathcal{Q}_{1}} \otimes V_{\mathcal{Q}_{2}}\right) 
\\
V_{\mathcal{Q}_{1}}^{\dagger} \otimes V_{\mathcal{Q}_{2}}^{\dagger} R^{(2)}_{1,2}(\ket{\psi}) 
& = & 
\frac{1}{2}\left( \ket{0}_{\mathcal{Q}_{1}} \ket{\psi_{1}}_{\mathcal{A}_{1}} + \ket{1}_{\mathcal{Q}_{1}}\ket{\psi_{1}^{*}}_{\mathcal{A}_{1}}\right) \otimes\\ \nonumber
& & \, \, \, \, \, \, \, \, \,
\left( \ket{0}_{\mathcal{Q}_{2}} \ket{\psi_{2}}_{\mathcal{A}_{2}} + \ket{1}_{\mathcal{Q}_{2}}\ket{\psi_{2}^{*}}_{\mathcal{A}_{2}}\right) 
\\
\bra{\psi} M_{\mathcal{A}}  \ket{\psi}
& = & 
R^{(2)}_{1,2}(\bra{\psi}) R^{(2)}_{1,2}(M_{\mathcal{A}}) R^{(2)}_{1,2}(\ket{\psi}) 
\end{eqnarray}
%¥
\end{lemma}
%¥

%
\begin{proof}
The first two properties follow directly from the definition and lemma~\ref{lemma:propertiesofR}.

For the last property, first we decompose $M$ into a sum of product operators from which we calculate the left side as
\begin{equation}
\sum_{k} \bra{\psi_{1}}M_{\mathcal{A}_{1},k} \ket{\psi_{1}}
	 \bra{\psi_{2}}M_{\mathcal{A}_{2},k} \ket{\psi_{2}}.
\end{equation}
Meanwhile, using the first two properties, the right hand side becomes
\begin{multline}
\frac{1}{4} \sum_{k} \bra{\psi_{1}}M_{\mathcal{A}_{1},k} \ket{\psi_{1}}
	 \bra{\psi_{2}}M_{\mathcal{A}_{2},k} \ket{\psi_{2}} +  
	 \bra{\psi_{1}^{*}}M^{*}_{\mathcal{A}_{1},k} \ket{\psi_{1}^{*}}
	 \bra{\psi_{2}}M_{\mathcal{A}_{2},k} \ket{\psi_{2}} + \\
	 \bra{\psi_{1}}M_{\mathcal{A}_{1},k} \ket{\psi_{1}}
	 \bra{\psi_{2}^{*}}M^{*}_{\mathcal{A}_{2},k} \ket{\psi_{2}^{*}} +
	 \bra{\psi_{1}^{*}}M^{*}_{\mathcal{A}_{1},k} \ket{\psi_{1}^{*}}
	 \bra{\psi_{2}^{*}}M^{*}_{\mathcal{A}_{2},k} \ket{\psi_{2}^{*}}.
\end{multline}
%¥
Since each $M_{\mathcal{A}_{j},k}$ is Hermitian and $\bra{\psi_{j}^{*}}M^{*}_{\mathcal{A}_{j},k} \ket{\psi_{j}^{*}} = \left(\bra{\psi_{j}}M_{\mathcal{A}_{j},k} \ket{\psi_{j}}\right)^{*}$, all the summands are in fact equal, and we obtain equality with the left hand side.

\end{proof}
%¥

%%%%%%%%%%%%%%%%%%%%%%%%%%%%%%%%%%%%%%%%%%%%%%
\subsection{Inverse transformation}
We now consider a type of inverse of $R(\cdot)$ in the case of a single system.  First, note that any real state $\ket{\psi^{\prime}}$ that has a qubit register $\mathcal{Q}$ can be written as
\begin{equation}
\ket{\psi^{\prime}} = \ket{0}_{\mathcal{Q}} \ket{\psi_{0}}_{\mathcal{A}} + \ket{1}_{\mathcal{Q}} \ket{\psi_{1}}_{\mathcal{A}}
\end{equation}
and the state $\ket{\psi} = \ket{\psi_{0}} + i \ket{\psi_{1}}$ satisfies $R(\ket{\psi}) = \ket{\psi^{\prime}}$.  Now, suppose that we have an efficient real circuit which outputs $\ket{\psi^{\prime}}$.  Does this imply that there is an efficient circuit which outputs $\ket{\psi}$?  Indeed it does.

\begin{lemma}\label{lem:rewind}
Let an $n$-qubit unitary $U$, implemented as a polynomial sized circuit, be given such that
\begin{equation}
U \ket{0}^{\otimes n} = 
\ket{\psi^{\prime}} = \ket{0} \ket{\psi_{0}} + \ket{1} \ket{\psi_{1}}
\end{equation}
where $\ket{\psi^{\prime}}$ is real.  Then there exists an $(n+1)$-qubit unitary $U^{\prime}$, implemented as a polynomial sized circuit, such that 

\begin{equation}
U^{\prime} \ket{0}^{\otimes n+1} = \ket{00}\left(\ket{\psi_{0}} + i\ket{\psi_{1}} \right) = \ket{\psi}
\end{equation}

\end{lemma}

\begin{proof} We use the ``quantum rewinding'' trick due to Watrous \cite{Watrous:2009:Zero-Knowledge-}.  First we use $V_{\mathcal{Q}_{1}}$, defined as in previous sections and applied to the first qubit, to form $W=V^{\dagger}_{\mathcal{Q}_{1}} U$, so 
\begin{equation}
W \ket{0}^{\otimes n} = \frac{1}{\sqrt{2}}\left(\ket{0} \ket{\psi} + \ket{1} \ket{\psi^{*}}\right).
\end{equation}
Next, let $\ket{\phi}$ be defined by
\begin{equation}
\ket{\phi} = W^{\dagger}\frac{1}{\sqrt{2}}\left(\ket{0} \ket{\psi} - \ket{1} \ket{\psi^{*}}\right).
\end{equation}
Note that $\braket{0^{\otimes n}}{\phi} = 0$.

Finally, let $N$ be given by:
\begin{equation}
N \ket{x} = 
\begin{cases}
\ket{x} & x = 0 \\
- \ket{x} & x \neq 0.
\end{cases}
\end{equation}
$N$ can be implemented by a polynomially sized circuit, for example using the standard construction for gates controlled on many qubits given in~\cite{Michael-A.-Nielsen:2000:Quantum-Computa}, figure 4.10.

\begin{figure}
\[
\Qcircuit { 
\lstick{\ket{0}} & \qw  & \targ & \ctrl{1} &  \ctrl{1} & \ctrl{1} & \gate{H} & \qw \\ 
\lstick{\ket{0}} & \multigate{1}{W} & \ctrl{-1} & \multigate{1}{W^{\dagger}}  & \multigate{1}{N} & \multigate{1}{W} & \qw & \qw \\
\lstick{\ket{0}^{\otimes (n-1)}} & \ghost{W} & \qw& \ghost{W^{\dagger}} & \ghost{N} & \ghost{W} & \qw & \qw \\
 }
\]	
\caption{Rewinding $U$}
\label{fig:rewindcircuit}
\end{figure}
Now we construct $U^{\prime}$ as in the circuit given in figure~\ref{fig:rewindcircuit}.  After the first two gates we obtain the state
\begin{equation}
\frac{1}{\sqrt{2}}\left( \ket{00}\ket{\psi}+ \ket{11}\ket{\psi^{*}} \right)
\end{equation}
and after the controlled $W^{\dagger}$, the state
\begin{equation}
\frac{1}{\sqrt{2}}\left( \ket{00}\ket{\psi}  + \frac{1}{\sqrt{2}}\ket{1}\left(\ket{0}^{\otimes n} - \ket{\phi}Ê\right)  \right).
\end{equation}
The controlled $N$ gate takes $\ket{\phi}$ to $-\ket{\phi}$, and we obtain
\begin{equation}
\frac{1}{\sqrt{2}}\left( \ket{\psi} \ket{00} + \frac{1}{\sqrt{2}}\ket{1}\left(\ket{0}^{\otimes n} + \ket{\phi}Ê\right)  \right).
\end{equation}
Finally, applying the final controlled $W$ and $H$, the state becomes
\begin{equation}
 \ket{0} \ket{0} \ket{\psi}.
\end{equation}

Recall that we can take a polynomial sized circuit for $W$ and construct a polynomial sized circuit for a controlled-$W$ and controlled-$W^{\dagger}$ using standard techniques \cite{Michael-A.-Nielsen:2000:Quantum-Computa}.  The circuit in figure~\ref{fig:rewindcircuit} is hence also polynomial sized.

\end{proof}

%%%%%%%%%%%%%%%%%%%%%%%%%%%%%%%%%%%%%%%%%%%%%%
\section{Complexity implications}
The real simulation for single systems and lemma \ref{lemma:propertiesofR} immediately imply that $\mathsf{BQP} = \mathsf{BQP}_{\mathbb{R}}$, $\mathsf{EQP} = \mathsf{EQP}_{\mathbb{R}}$ and $\mathsf{QCMA} = \mathsf{QCMA}_{\mathbb{R}}$ (interactive proofs with a single classical message and a quantum verifier).  We show that the analogous results are also true for $\mathsf{QMA}$, $\mathsf{QMA}(2)$, $\mathsf{QIP}$, $\mathsf{MQIP}$ and $\mathsf{QSZK}$.

%%%%%%%%%%%%%%%%%%%%%%%%%
\subsection{Imaginary measurements}
Before we begin, we first discuss a proof technique which we will make frequent use of, which is to imagine that certain measurements have been performed.  In particular, we will frequently make use of equation~\eqref{eq:vbasis} or the analogous equations for the multipartite and separable real simulations.  These equations show that the simulation operators all have a particular form which commutes with the operator $iV_{\mathcal{Q}}ZV^{\dagger}_{\mathcal{Q}} = Y_{\mathcal{Q}}$.  If, for example, the verifier is a real simulation the all its operators commute with $Y_{\mathcal{Q}}$ so we can in principle measure $Y_{\mathcal{Q}}$ at any time without affecting the outcome of the calculation.  Although invisible from the perspective of the outcome of the computation, these measurements impose a structure on the state which we can take advantage of.  In particular, due to the form in equation~\eqref{eq:vbasis}, the outcome of a $Y_{\mathcal{Q}}$ measurement dictates whether a simulation operator behaves like the original operator, or its complex conjugate.

We use this technique as follows.  We begin with a calculation $C$ which accepts with probability $p$.  We note that $C$ commutes with a measurement $Y$ so we imagine a new calculation $C^{\prime}$ in which the measurement $Y$ is performed at some point.  Since $Y$ commutes with $C$, $C^{\prime}$ also accepts with probability $p$.  Now we use the additional structure imposed on $C^{\prime}$ to show that $C^{\prime}$ accepts with probability less than $s$ (or greater than $c$, depending on what we are proving), and hence $p \leq s$.  Finally, we see that $C$ accepts with probability less than $s$.

This basic argument will be used repeatedly for the different complexity classes.

%%%%%%%%%%%%%%%%%%%%%%%%%
\subsection{$\mathsf{QMIP}$}\label{sec:reversecontainment}

Interestingly, we must first show that, by limiting to real numbers, we do not actually increase the power of a complexity class.  Imagine a real verifier engaging in a protocol with a real prover.  If we then allow the prover access to a complex Hilbert space, it has access to a larger set of operations and may be able to cheat against the real verifier with higher probability.  Hence  that there could be problems in $\mathsf{QMIP}[m,k]_{\mathbb{R}}$ that are not in $\mathsf{QMIP}[m,k]$.  However, as we shall see, this is not the case and in fact $\mathsf{QMIP}[m,k]_{\mathbb{R}} \subseteq \mathsf{QMIP}[m,k]$.  By substituting in suitable values of $m$ and $k$ we obtain the analogous result for $\mathsf{QIP}$, and $\mathsf{QMA}$.  We leave $\mathsf{QMA}(2)$, $\mathsf{QMIP}_{ne}$ and $\mathsf{QSZK}$ as special cases.

\begin{lemma}
\begin{equation}
\mathsf{QMIP}[m,k]_{\mathbb{R}} \subseteq \mathsf{QMIP}[m,k]
\end{equation}
\end{lemma}
%¥

\begin{proof}

Suppose that we have a verifier $A(x)$ for some language $L$ in $\mathsf{QMIP}[m,k]_{\mathbb{R}}$.  Now we lift $A(x)$ to a complex Hilbert space.  Clearly all states in the real Hilbert space can also be lifted to the complex Hilbert space, so the soundness and completeness cannot go down in the complex setting.  Now let a set of complex provers $P_{j}(x)$ be given that cause $A(x)$ to accept with probability $p$.  We can construct a different set of provers $P_{j}^{\prime}(x)$ and verifier $A^{\prime}(x)$ by using the multi-party real simulation from section \ref{sec:multipartite}.  However, since each of $A(x)$'s operations $M$ is real, the image $R^{(m+1)}_{m+1}(M)$ has the form $I_{\mathcal{Q}_{m+1}} \otimes M$.  In fact, the new verifier does not need $\mathcal{Q}_{1}$ at all and we can take $A^{\prime}(x) = A(x)$.

Now the provers in $P_{j}^{\prime}(x)$ are all real.  They must begin with an entangled state, but this is allowed in the definition of $\mathsf{QMIP}[m,k]$.  As mentioned above, $A(x)$ does not need access to the entangled state on $\mathcal{Q}_{1}$ for the simulation to work.  Now when the provers $P_{j}^{\prime}(x)$ interact with $A(x)$ they cause it to accept with probability $p$, i.e.Êwith the same probability as the complex provers $P_{j}(x)$.  This shows that the completeness and soundness cannot go up when we lift $A(x)$ to the complex setting.  Thus $\mathsf{QMIP}[m,k]_{\mathbb{R}} \subseteq \mathsf{QMIP}[m,k]$.

\end{proof}

\begin{lemma}
\begin{equation}
\mathsf{QMIP}[m,k] \subseteq \mathsf{QMIP}[m,k]_{\mathbb{R}}
\end{equation}
\end{lemma}
%¥

\begin{proof}
Here we have two tasks:  show how to use the real simulation from section~\ref{sec:multipartite} to keep the same completeness, and prove that the soundness is also preserved.  As in the previous section by substituting in suitable values of $m$ and $k$ we obtain the analogous result for $\mathsf{QIP}$, and $\mathsf{QMA}$ and leave $\mathsf{QMA}(2)$, $\mathsf{QMIP}_{ne}$ and $\mathsf{QSZK}$ as special cases.

We begin with a complex verifier $A(x)$ and provers $P_{j}(x)$ for some protocol.  We can apply the real simulation to find real verifier $A^{\prime}(x)$ and provers $P_{j}^{\prime}(x)$.  We need to distribute the entangled state in registers $\mathcal{Q}_{1} \dots \mathcal{Q}_{m+1}$ to all the parties.  In the case where $k$ is odd, so that the provers go first, they may begin with the state already shared.  Prover 1 begins with registers $\mathcal{Q}_{1}$ and $\mathcal{Q}_{m+1}$, and sends $\mathcal{Q}_{m+1}$ to the verifier in the first message.  If $k$ is even, then the verifier prepares the all the $\mathcal{Q}$ registers and sends $\mathcal{Q}_{j}$ to prover $j$ in the first message, keeping $\mathcal{Q}_{m+1}$ for himself.  In this way we have constructed a real protocol with the same completeness as the original complex protocol.

We now show that the above construction preserves soundness.  For this argument we consider everything over complex numbers.  Note that this only gives more power to the provers, and can only increase the soundness.  Recall from section~\ref{sec:multipartite} that there exists a local change of basis $V_{\mathcal{Q}_{m+1}}$ such that an operator $M$ applied by the verifier becomes
\begin{equation}
V_{\mathcal{Q}_{m+1}}^{\dagger}R_{\mathcal{Q}_{m+1}}(M)V_{\mathcal{Q}_{m+1}} = \proj{0}{0}_{\mathcal{Q}_{m+1}}\otimes M + \proj{1}{1}_{\mathcal{Q}_{m+1}} M^{*}.
\end{equation}
Hence in the real protocol all operations commute with $iV_{\mathcal{Q}_{m+1}}ZV_{\mathcal{Q}_{m+1}}^{\dagger} = Y_{\mathcal{Q}_{m+1}}$.  There is no difference, then, if we measure $Y_{\mathcal{Q}_{m+1}}$ so we may imagine that the verifier does so before any other operations.

Now we come to the crux of the argument.  As above, after changing basis by $V_{\mathcal{Q}_{m+1}}$ each operation $M^{\prime}$ of the verifier $A^{\prime}(x)$ looks like a controlled operation, with the control on $\mathcal{Q}_{m+1}$.  So if the $Y_{\mathcal{Q}_{m+1}}$ measurement outcome was ``0'' then the verifier performs $M$ as in the original complex protocol, and the overall behaviour of $A^{\prime}(x)$ is the same as $A(x)$.  If the outcome was ``1'' then the verifier performs $M^{*}$ and the behaviour is that of $A^{*}(x)$, the complex conjugate of $A(x)$.  Clearly in both cases the soundness is the same as for the original verifier $A(x)$ and we have merely taken a convex combination through the $Y_{\mathcal{Q}_{m+1}}$ measurement.  Hence in the real protocol the soundness is identical to the original complex verifier $A(x)$.  
\end{proof}

One might wonder whether the multiparty real simulation is necessary in the case of $\mathsf{QIP}$.  Since the verifier and prover take turns they could communicate a single ``extra'' qubit back and forth.  In fact, this is not sufficient.  In~\ref{sec:singlepartynotsuff} we give an explicit example where the soundness of such a construction is not the same as in the original protocol.

%%%%%%%%%%%%%%%%%%%%%%%%%
\subsection{ $\mathsf{QMA}(k)$}\label{sec:qma}
We begin with a language $L \in \mathsf{QMA}(k)$.  For input $x \in L$ we may model the verifier $A(x)$ as the positive semi-definite operator (POVM element) such that
\begin{equation}
P(\text{ACCEPT}) = \bra{\psi_{1}} \dots \bra{\psi_{k}} A(x) \ket{\psi_{1}} \dots \ket{\psi_{k}}
\end{equation}
We recall a recent result of Harrow and Montanaro \cite{Harrow:2010:An-efficient-te} wherein the authors prove that $\mathsf{QMA}(k) = \mathsf{QMA}^{\text{SEP}}(2)$.  Here $\mathsf{QMA}^{\text{SEP}}(2)$ is $\mathsf{QMA}(2)$ with the additional restriction that $A(x)$ is separable across the message registers from the two provers.  Harrow and Montanaro give an explict contstruction for two provers and a separable verifier from $k$ provers and an arbitrary verifier.  For our purposes we also need the equality $\mathsf{QMA}_{\mathbb{R}}(k) = \mathsf{QMA}_{\mathbb{R}}^{\text{SEP}}(2)$, which can be seen by noting that Harrow and Montanaro's construction does not introduce any complex numbers, hence real verifiers remain real.  From this, it suffices to show  $\mathsf{QMA^{\text{SEP}}(2)} = \mathsf{QMA(2)}_{\mathbb{R}}^{\text{SEP}}$.

Recall from the definition of $\mathsf{QMA}(k)$ that the provers must send unentangled states.  Thus we are in the situation of an unentangled state and separable measurement, described in section~\ref{sec:separable}.  We proceed in two steps, showing  $\mathsf{QMA^{\text{SEP}}(2)}$ and $\mathsf{QMA(2)}_{\mathbb{R}}^{\text{SEP}}$ contain eachother.

\begin{lemma}
\begin{equation}
\mathsf{QMA}(2)_{\mathbb{R}}^{\text{SEP}} \subseteq \mathsf{QMA}(2)^{\text{SEP}}
\end{equation}

\end{lemma}
%¥
\begin{proof}
Let $A(x) = \sum_{k} A_{k}(x) \otimes B_{k}(x)$ be a separable verifier for a language $L \in \mathsf{QMA}(2)_{\mathbb{R}}^{\text{SEP}}$.  The $A$ and $B$ operators act on the two message registers from the two provers, respectively.  For some state $\ket{\psi_{1}} \ket{\psi_{2}}$ we lift $A(x)$ and $\ket{\psi_{1}} \ket{\psi_{2}}$ to a complex Hilbert space and note that the probability of acceptance is the same.  Hence the soundness and completeness cannot go down in the complex setting. 

Now we prove that soundness and completeness cannot go up. Suppose $A(x)$ accepts a complex state $\rho_{A} \otimes \rho_{B}$ (where $A$ and $B$ registers come from the first and second provers, respectively) with probability $p$.  We may write $\rho_{A}$ and $\rho_{B}$ as real and imaginary parts, obtaining
\begin{equation}
p = \sum_{k} \tr{A_{k}(x) \rho_{A}^{R}} \tr{B_{k}(x) \rho_{B}^{R}} 
+ i\tr{A_{k}(x) \rho_{A}^{R}} \tr{B_{k}(x) \rho_{B}^{I}} 
\end{equation}
\begin{equation*}
+ i\tr{A_{k}(x) \rho_{A}^{I}} \tr{B_{k}(x) \rho_{B}^{R}}
-\tr{A_{k}(x) \rho_{A}^{I}} \tr{B_{k}(x) \rho_{B}^{I}}
\end{equation*}
where $\rho^{R}_{A}$ and $\rho_{B}^{R}$ are real symmetric and $\rho^{I}_{A}$ and $\rho^{I}_{B}$ are real anti-symmetric  ($(\rho^{I}_{A})_{ab} = -(\rho^{I}_{A})_{ba}$).  It is easy to show that $\tr{MN} = 0$ whenever $M$ is symmetric and $N$ is anti-symmetric, so the last three terms in the above summation are all zero, and
\begin{equation}
p = \sum_{k} \tr{A_{k}(x) \rho_{A}^{R}} \tr{B_{k}(x) \rho_{B}^{R}} =  \sum_{k} \tr{A_{k}(x)  \otimes B_{k}(x)\rho_{A}^{R}\otimes \rho_{B}^{R}}
\end{equation}
We can then decompose $\rho_{A}^{R}$ and $\rho_{B}^{R}$ into mixtures of pure states and, by convexity, obtain at least one real pure state $\ket{\phi}_{A}\ket{\phi}_{B}$ which the verifier will accept with probability at least $p$.  Hence the soundness and completeness cannot increase when lifting the real verifier into the complex setting, and $\mathsf{QMA}_{\mathbb{R}}(2)^{\text{SEP}} \subseteq \mathsf{QMA}(2)^{\text{SEP}}$.
\end{proof}

\begin{lemma}
\begin{equation}
\mathsf{QMA}(2)_{\mathbb{R}}^{\text{SEP}} \supseteq \mathsf{QMA}(2)^{\text{SEP}}
\end{equation}

\end{lemma}
%¥
\begin{proof}
Let $L$ be a laungage in $\mathsf{QMA}(k)^{\text{SEP}}$ with separable verifier $A(x)$, which has completeness $c$ and soundness $s$.  Suppose there exists a valid proof $\ket{\psi_{1}}\ket{\psi_{2}}$.  Applying the simulation in section \ref{sec:separable} prover 1 can send $R^{(2)}_{1}(\ket{\psi_{1}})$ while prover 2 can send $R^{(2)}_{2}(\ket{\psi_{2}})$.  Meanwhile, the verifier becomes $R^{(2)}_{1,2}(A(x))$ since it is separable.  We then have
\begin{equation}
R^{(2)}_{1}(\bra{\psi_{1}})R^{(2)}_{2}(\bra{\psi_{2}}) R^{(2)}_{1,2}(A(x)) R_{1}(\ket{\psi_{1}})R^{(2)}_{2}(\ket{\psi_{2}}) = \bra{\psi_{1}}\bra{\psi_{2}}A(x) \ket{\psi_{1}} \ket{\psi_{2}} \geq c
\end{equation}
so the completeness of $R^{(2)}_{1,2}(A(x))$ is again $c$.

To analyze soundness we use an argument analogous to the $\mathsf{QMIP}$ case.  Suppose that some $x$ is not in $L$, so $\bra{\psi}A(x)\ket{\psi} \leq s$ for all $\ket{\psi}$.  By lemma~\ref{lemma:propertiesofRmsep} 
\begin{equation}
V_{\mathcal{Q}_{1}}^{\dagger}Ê\otimes V_{\mathcal{Q}_{2}}^{\dagger} R^{(2)}_{1,2}(A(x)) V_{\mathcal{Q}_{1}} \otimes V_{\mathcal{Q}_{2}} = \sum_{z \in \{0,1\}^{2}} \proj{z}{z}_{\mathcal{Q}_{1}\mathcal{Q}_{2}} \otimes A(x)^{*_{z}} .
\end{equation}
Thus measuring $iV_{\mathcal{Q}_{1}}ZV^{\dagger}_{\mathcal{Q}_{1}} = Y_{\mathcal{Q}_{1}}$ commutes with $R^{(2)}_{1,2}(A(x))$ and we may imagine that this measurement happens, and in particular that prover 1 measures before sending the state.  Similarly, we may imagine that prover 2 measures $\mathcal{Q}_{2}$ in the basis $Y_{\mathcal{Q}_{2}}$.  Equivalently, the provers might as well have sent a mixture of $V_{\mathcal{Q}_{1}}\ket{j}_{\mathcal{Q}_{1}}\ket{\psi_{j}} V_{\mathcal{Q}_{2}}\ket{k}_{\mathcal{Q}_{2}}\ket{\phi_{k}}$ for $j,k \in \{0,1\}$ and for some complex $\ket{\psi_{j}}$ and $\ket{\phi_{k}}$.  Note that the resulting mixture need not be real, but by considering complex states we only increase the power of the cheating provers.  The probability of accepting is then a convex combination of the form
\begin{equation}
\sum_{z \in \{0,1\}^{2}} p_{z} \bra{\psi_{z_{1}}} \bra{\phi_{z_{2}}} A^{*_{z}}(x) \ket{\psi_{z_{1}}} \ket{\phi_{z_{2}}}.
\end{equation}
By lemma~\ref{lemma:separableconjugation} and the soundness of $A(x)$, each term $\bra{\psi_{z_{1}}} \bra{\phi_{z_{2}}} A^{*_{z}}(x) \ket{\psi_{z_{1}}} \ket{\phi_{z_{2}}}$ is at most $s$ and hence by convexity the overall acceptance probability is at most $s$.  Thus $R^{(2)}_{1,2}(A(x))$ is a verifier with the same completeness and soundness as $A(x)$ and we have proved that $\mathsf{QMA^{\text{SEP}}(2)} \subseteq \mathsf{QMA(2)}_{\mathbb{R}}^{\text{SEP}}$.
\end{proof}

We should point out that, unlike the proof for $\mathsf{QMIP}$, here we are not allowed to simulate just any protocol for a $\mathsf{QMA}(k)$ problem.  It must first be transformed into $\mathsf{QMA^{SEP}}(2)$ protocol before the mapping to a real protocol is applied.

%%%%%%%%%%%%%%%%%%%%%%%%%%%%%%%%%%%%%%%%%%
\subsection{ $\mathsf{QMIP_{ne}}_{\mathbb{R}}$}
For the case of non-entangled provers we need to make small adjustments which, in some cases, increase the number of messages by 1.

\begin{lemma}
\begin{equation}
\mathsf{QMIP_{ne}}_{\mathbb{R}}[m,k] \subseteq \mathsf{QMIP_{ne}}[m,k+1] 
\end{equation}
\end{lemma}
%¥
\begin{proof}
Here the argument follows that in section~\ref{sec:reversecontainment}.  However, there is a problem which arises when we find the multi-party simulation of a set of cheating provers.  Since entanglement is required, and the provers do not start with any, it must be provided by the verifier.  Thus we add an additional message at the start of the protocol where the verifier prepares $\mathcal{Q}_{1} \dots \mathcal{Q}_{m+1}$ in the necessary state and sends $\mathcal{Q}_{j}$ to prover $j$.  This allows the simulation to proceed and increases the number of messages by at most 1.

We have potentially created a problem by handing over entanglement to the provers.  However, the verifier keeps $\mathcal{Q}_{m+1}$ so that the state that the provers have, once $\mathcal{Q}_{m+1}$ traced out, is just a shared random bit.  In particular, the provers share a mixture of $V_{\mathcal{Q}_{1}} \otimes \dots \otimes V_{\mathcal{Q}_{m}} \frac{1}{\sqrt{2}}\ket{00 \dots 0}_{\mathcal{Q}_{1} \dots \mathcal{Q}_{m}}$ and $V_{\mathcal{Q}_{1}} \otimes \dots \otimes V_{\mathcal{Q}_{m}} \frac{1}{\sqrt{2}}\ket{11 \dots 1}_{\mathcal{Q}_{1} \dots \mathcal{Q}_{m}} $.  This shared random bit can only allow the provers to take a convex combination of strategies, not increase the maximum cheating probability.  Hence giving the provers the entanglement in this way does not give any additional power to them.
\end{proof}
\begin{lemma}
\begin{equation}
\mathsf{QMIP_{ne}}[m,k] \subseteq \mathsf{QMIP_{ne}}_{\mathbb{R}}[m,k+1]
\end{equation}

\end{lemma}

\begin{proof}
The argument is the same as in the case when entanglement is allowed, except that we need to establish the necessary entanglement between provers.  If the verifier goes first, then it prepares $\mathcal{Q}_{1} \dots \mathcal{Q}_{m+1}$ in the necessary state sends $\mathcal{Q}_{j}$ to prover $j$ along with the first message.  If not, then we insert a new message at the beginning of the protocol that does just this, and increases the number of messages by 1.  
\end{proof}

%%%%%%%%%%%%%%%%%%%%%%%%%
\subsection{ $\mathsf{QSZK}$}
$\mathsf{QSZK}$ is essentially $\mathsf{QIP}$ with the additional restriction that, assuming an honest prover, any state the verifier can create while interacting with the prover must be close in trace distance to the output of some efficient quantum circuit.  We need only verify this additional condition since completeness and soundness are dealt with in the $\mathsf{QIP}$ case.

First, we need a lemma about the real simulation:
\begin{lemma}\label{lem:simdistance}
Let $\ket{\psi}_{\mathcal{A} \mathcal{B}}$ and $\ket{\phi}_{\mathcal{A}\mathcal{B}}$ be a bipartite complex states and $\ket{\psi^{\prime}}_{\mathcal{A}\mathcal{B}\mathcal{Q}} = R(\ket{\psi})$, $\ket{\phi^{\prime}}_{\mathcal{A}\mathcal{B}\mathcal{Q}} = R(\ket{\phi})$ be the real simulation states.  Let 
\begin{eqnarray*}
\rho &=&\tr[\mathcal{A}]{\proj{\psi}{\psi}}\\
\sigma &=& \tr[\mathcal{A}]{\proj{\phi}{\phi}} \\
\rho^{\prime} &=&\tr[\mathcal{A}\mathcal{Q}]{\proj{\psi^{\prime}}{\psi^{\prime}}}\\
\sigma^{\prime} &=& \tr[\mathcal{A}\mathcal{Q}]{\proj{\phi^{\prime}}{\phi^{\prime}}}
\end{eqnarray*}
Then
\begin{equation}
\norm{\rho - \sigma}_{1} \leq \norm{\rho^{\prime} - \sigma^{\prime}}_{1}
\end{equation}

\end{lemma}
\begin{proof}
First we transform the real states by $V^{\dagger}_{\mathcal{Q}}$, which does not change the trace distance.  From now on we implicitly assume this change of basis has occurred.  After this change of basis we find
\begin{equation}
\rho^{\prime} = \frac{1}{2}\left(
	\begin{matrix}
	\rho_{00} & \rho_{01} \\
	\rho_{10} & \rho_{11}
	\end{matrix}
\right)
\end{equation}
with $\rho_{00} = \rho_{11}^{*} = \rho$ since $\ket{\psi^{\prime}} = \frac{1}{\sqrt{2}}\left(\ket{0}\ket{\psi} + \ket{1}\ket{\psi^{*}}\right)$.  Now consider the superoperator $\Phi(\cdot)$ which measures the $\mathcal{Q}$ register in the $Z_{\mathcal{Q}}$ eigenbasis and records the result back into the $\mathcal{Q}$ register.  Then $\Phi(\cdot)$ just zeros out the off-diagonal blocks, and
\begin{equation}
\Phi(\rho^{\prime}) = \frac{1}{2}\left(
	\begin{matrix}
	\rho & 0 \\
	0 & \rho^{*}
	\end{matrix}
\right)
\end{equation}
and analogously
\begin{equation}
\Phi(\sigma^{\prime}) = \frac{1}{2}\left(
	\begin{matrix}
	\sigma & 0 \\
	0 & \sigma^{*}
	\end{matrix}
\right)
\end{equation}
Next we make two observations.  First, since superoperators cannot increase the trace distance $\norm{\Phi(\rho^{\prime}) - \Phi(\sigma^{\prime})}_{1} \leq \norm{\rho^{\prime}- \sigma^{\prime}}_{1}$.  Second, since we can diagonalize the two  blocks on the diagonal separately, and complex conjugation does not change the norm, $\norm{\Phi(\rho^{\prime}) - \Phi(\sigma^{\prime})}_{1} = \norm{\rho - \sigma}_{1}$.  Hence 
$\norm{\rho - \sigma}_{1} \leq \norm{\rho^{\prime} - \sigma^{\prime}}_{1}.$
\end{proof}

\begin{lemma}
\begin{equation}
\mathsf{QSZK}_{\mathbb{R}} \subseteq \mathsf{QSZK}
\end{equation}

\end{lemma}
\begin{proof}
Let $A(x)$ be a real verifier for a language in $\mathsf{QSZK}_{\mathbb{R}}$, which we lift to a complex Hilbert space.  Note that the completeness and soundness do not change, as shown in section~\ref{sec:reversecontainment}, so we just need to show that a real cheating verifier cannot trick a honest real prover $P$ into revealing too much information. 

Let $W$ be some polynomially sized complex circuit which interacts with $P$ and which outputs state $\ket{\psi}_{\mathcal{A}\mathcal{B}}$ on some output qubits $\mathcal{B}$ and non-output qubits $\mathcal{A}$.  We assume that $\mathcal{A}$ includes any part of the state that $P$ keeps.  Then consider $\rho = \tr[\mathcal{A}]{\proj{\psi}{\psi}}$.  We can form circuit $W^{\prime}$ and state $\ket{\psi^{\prime}}_{\mathcal{A}\mathcal{B}\mathcal{Q}}$ by applying $R(\cdot)$.  Note that $P$ is not changed (besides adding an identity on $\mathcal{Q}$) since it is real.  The output state of the real circuit is $\rho^{\prime} = \tr[\mathcal{A}]{\proj{\psi^{\prime}}{\psi^{\prime}}}$.  

From the definition of $\mathsf{QSZK}_{\mathbb{R}}$, there must be some state $\sigma^{\prime}$ such that $\norm{\rho^{\prime} - \sigma^{\prime}}_{1} \leq c$, and $\sigma^{\prime} = \tr[\mathcal{A}]{\proj{\phi^{\prime}}{\phi^{\prime}}}$ where $\ket{\phi^{\prime}}_{\mathcal{A}\mathcal{B}}$ is the output of some polynomially sized circuit $U^{\prime}$, not involving $P$. By lemma~\ref{lem:rewind} there is a polynomially sized circuit $U$ (which also does not interact with $P$) which outputs complex state $\ket{\phi}$ such that $R(\ket{\phi}) = \ket{\phi^{\prime}}$.  After tracing out non-output qubits the state becomes $\sigma = \tr[A]{\proj{\phi}{\phi}}$.  Finally, by lemma~\ref{lem:simdistance}
\begin{equation}
\norm{\rho - \sigma}_{1} \leq \norm{\rho^{\prime} - \sigma^{\prime}}_{1} \leq c.
\end{equation}
Hence we have shown that any state that a complex cheating verifier can create by interacting with $P$ is close to a state that can be efficiently created not using $P$.

\end{proof}

\begin{lemma}
\begin{equation}
\mathsf{QSZK}_{\mathbb{R}} \supseteq \mathsf{QSZK}
\end{equation}
\end{lemma}
%¥

\begin{proof}
{\bf Structure of the prover}
Let $A(x)$ be the verifier for some language $LÊ\in \mathsf{QSZK}$, and let $P(x)$ be the corresponding honest prover.  We use the same construction as for $\mathsf{QIP} \subseteq \mathsf{QIP}_{\mathbb{R}}$, from which we obtain a real verifier $A^{\prime}(x)$ and a real prover $P^{\prime}$.  We also learn that for the real verifier the completeness and soundness are the same as for the complex verifier.

The remaining property to show is that for any real polynomially sized real circuit $W^{\prime}$, which interacts with $P^{\prime}$ and outputs a state $\rho^{\prime}$ there is another real, polynomially sized circuit $U^{\prime}$ which outputs a state $\sigma^{\prime}$ for which $\norm{\rho^{\prime} - \sigma^{\prime}}_{1} \leq c$.  We will first construct a complex $W$ which interacts with $P$ to produce a state $\rho$ and use the property of $\mathsf{QSZK}$ that there exists a polynomially sized circuit $U$ which outputs $\sigma$ such that $\norm{\rho - \sigma} \leq c$.  There are two cases to argue, when $P$ sends the first message, and when $W$ sends the first message.  From this we construct a real $U^{\prime}$ that outputs the required $\sigma^{\prime}$.

Before we analyze the two cases, we first consider the structure of $P^{\prime}$.  To this end, we apply $V^{\dagger}_{\mathcal{Q}_{1}}$ to each of $P^{\prime}$'s operations, after which they have the form
\begin{equation}
V^{\dagger}_{\mathcal{Q}_{1}} M^{\prime} V_{\mathcal{Q}_{1}} = \proj{0}{0}_{\mathcal{Q}_{1}} \otimes M + \proj{1}{1}_{\mathcal{Q}_{1}} \otimes M^{*}
\end{equation}
where $M$ is $P$'s original operation.  Measuring in the $iV_{\mathcal{Q}_{1}}Z_{\mathcal{Q}_{1}}V^{\dagger}_{\mathcal{Q}_{1}} = Y_{\mathcal{Q}_{1}}$ eigenbasis thus commutes with each of $P^{\prime}$'s operations.  In the following we may thus imagine that $\mathcal{Q}_{1}$ has been measured in this basis.

Let us for a moment not trace out $\mathcal{Q}_{1}$ and keep it along with the verifier's output qubits.  Conjugating this state by $V^{\dagger}_{\mathcal{Q}_{1}}$ we obtain a state
\begin{equation}
\rho^{\prime\prime} = 
\left(
	\begin{matrix}
	\rho_{0} & 0 \\
	0 & \rho_{1}Ê\\
	\end{matrix}
\right)
\end{equation}
where we have taken the liberty of measuring $\mathcal{Q}_{1}$ in the $Y_{\mathcal{Q}_{1}}$ basis.  (To be clear, the verifier's output state is $ \rho^{\prime} = \rho_{0} + \rho_{1}$.)  The state was real before we conjugate by $V_{\mathcal{Q}_{1}}$, i.e. 
\begin{equation}
V^{\dagger}_{\mathcal{Q}_{1}}\left(
	\begin{matrix}
	\rho_{0} & 0 \\
	0 & \rho_{1}Ê\\
	\end{matrix}
\right)
V_{\mathcal{Q}_{1}}
= 
\frac{1}{2}
\left(
	\begin{matrix}
	\rho_{0} + \rho_{1} & -i(\rho_{0} - \rho_{1})Ê\\
	i(\rho_{0} - \rho_{1}) & \rho_{0} + \rho_{1} \\
	\end{matrix}
\right)
\end{equation}
is real.  From this we deduce that $\rho_{0} = \rho_{1}^{*}$ which we will rename $\rho$.

Now let us construct a cheating complex verifier that outputs $\rho$ when interacting with $P$.  To this end we consider the interaction between $P^{\prime}$ and the cheating real verifier, viewed in the $V$ basis on $\mathcal{Q}_{1}$.  We consider two cases.  

{\bf Case 1: prover goes first}
In the first case the Prover sends the first message.  In this case $P^{\prime}$ would create an entangled pair of qubits $\mathcal{Q}_{1}\mathcal{Q}_{2}$ and send $\mathcal{Q}_{2}$ to the verifier.  Since measuring $\mathcal{Q}_{1}$ in the $Y_{\mathcal{Q}_{1}}$ basis commutes with all operations lets suppose that this happens.  In the case the outcome was $0$ the verifier's qubit $\mathcal{Q}_{2}$ would be in the state $\ket{+i}_{\mathcal{Q}_{2}}$ and $P^{\prime}$'s operations correspond to those of $P$.  For our complex cheating verifier, then, we simulate this process by preparing a qubit in the state $\ket{+i}_{\mathcal{Q}_{2}}$ and proceeding just as the real cheating verifier would, interacting with $P$.  This produces the state $\rho$.

{\bf Case 2: verifier goes first}
In the second case the verifier sends the first message to the prover.  Let $\ket{\psi^{\prime}} = \ket{0}_{\mathcal{Q}_{1}}\ket{\psi_{0}} + \ket{1}_{\mathcal{Q}_{1}} \ket{\psi_{1}}$ be the cheating real verifier's state immediately before sending the first message, where the qubit $\mathcal{Q}_{1}$ is sent to the prover for the simulation.  Changing basis by $V_{\mathcal{Q}_{1}}$ this becomes
\begin{equation}
V_{\mathcal{Q}_{1}} \ket{\psi^{\prime}} = \ket{0}_{\mathcal{Q}_{1}}\ket{\psi} + \ket{1}_{\mathcal{Q}_{1}} \ket{\psi^{*}}
\end{equation}
where $\ket{\psi} = \ket{\psi_{0}} + i \ket{\psi_{1}}$.  Note that by lemma~\ref{lem:rewind} there is an efficient method of preparing $\ket{\psi}$.  Now suppose that we measure $\mathcal{Q}_{1}$ in the $Y_{\mathcal{Q}_{1}}$ basis.  If the outcome is 0 then the real prover behaves as $P$ and the verifier's state is $\ket{\psi}$.  For the cheating complex prover, then, we simply prepare $\ket{\psi}$ as the verifier's initial state (dropping the unneeded qubit $\mathcal{Q}_{1}$) and proceed as in the real verifier, interacting with $P$.  The output state is then $\rho$.

We have shown that in all cases there is a cheating complex verifier that interacts with $P$ to produce $\rho$.  By complex conjugating, we obtain a cheating complex verifier that interacts with $P^{*}$ to produce $\rho^{*}$.  

We now use the property of the complex problem in $\mathsf{QSZK}$ that there is a polynomially sized circuit $U$ which outputs a state $\sigma$ so that $\norm{\rho - \sigma} \leq c$.  We construct a polynomially sized real circuit $U^{\prime}$ by applying the real simulation $R(\cdot)$ to each gate in the circuit.  Hence a gate $M$ is replaced with $M^{\prime} = V_{\mathcal{Q}_{2}}^{\dagger} \left(\proj{0}{0}Ê\otimes MÊ+ \proj{1}{1} \otimes M^{*}\right)$.  The qubit $\mathcal{Q}_{2}$ is prepared in the state $I/2$.  The output of this circuit is then $\sigma^{\prime\prime}$, and when conjugated by $V_{\mathcal{Q}_{2}}$ it is
$
V_{\mathcal{Q}_{2}}\sigma^{\prime\prime} V^{\dagger}_{\mathcal{Q}_{2}}= 
\frac{1}{2}\left(
	\begin{matrix}
	\sigma & 0 \\
	0 & \sigma^{*}
	\end{matrix}
\right)
$ 
with $\norm{\rho^{\prime\prime} - \sigma^{\prime\prime}} \leq c$.  Tracing out $\mathcal{Q}_{2}$ gives a state $\sigma^{\prime}$ such that $\norm{\rho^{\prime} - \sigma^{\prime}} \leq c$ since the trace norm is decreasing under partial trace.  Hence there is a real circuit which outputs a state that is close to $\rho^{\prime}$.

\end{proof}

%%%%%%%%%%%%%%%%%%%%%%%%%%%%%%%%%%%%%%%%%%%%%%
\section{Conclusions}
We have demonstrated that a wide variety of important quantum complexity classes are unchanged when the quantum operations are restricted to be over a real Hilbert space.  The arguments that we use are quite general and could be applied to other complexity classes as well.

{\bf Acknoledgements} This work is funded by the Centre for Quantum Technologies, which is funded by the Singapore Ministry of Education and the Singapore National Research Foundation.  Thanks to Bill Rosgen for helpful discussions.

\bibliographystyle{halphads}
\bibliography{Global_Bibliography}

\appendix

%%%%%%%%%%%%%%%%%%%%%%%%%%%%%%%%%%%%%%
\section{The single party real simulation is not sufficient to show $\mathsf{QIP}(k) \subseteq \mathsf{QIP}_{\mathbb{R}}(k)$.}\label{sec:singlepartynotsuff}

Suppose we make the following argument: the prover and verifier take turns performing operations, so we really only need one ``extra'' qubit, and they can pass it back and forth along with the messages.  We can thus transform the states and all operations according to $R(\cdot)$.  Clearly the completeness is the same, but what about soundness?  Here we give an example problem with an instance not in the language where in the original protocol the verifier accepts with probability arbitrarily close to 0, but for the real simulation with only one ``extra'' qubit there is a cheating prover that forces the verifier to always accept.

The problem we consider is quantum state distinguishability.  This is a complete problem for the class $\mathsf{QSZK}$ \cite{Watrous:2002:Limits-on-the-p}.  We are given two efficient quantum circuits $Q_{0}$ and $Q_{1}$ that produce outputs $\rho_{1}$ and $\rho_{2}$ after tracing out non-output qubits,  and the promise that either
\begin{equation}
\norm{\rho_{1} - \rho_{2}}_{1} \leq \alpha
\end{equation}
or
\begin{equation}
\norm{\rho_{1} - \rho_{2}}_{1} \geq \beta
\end{equation}
with $\alpha, \beta \in [0,1]$ and $\alpha < \beta^{2}$.  The problem is to determine which of these is true.  The basic solution to this problem is to prepare one of $\rho_{0}$ or $\rho_{1}$ at random and send it to the verifier.  The prover attempts to decide which state was sent, and sends a guess to the verifier.  The verifier the checks whether the prover was correct.  If the prover is very often correct over many instances of this game, then the states must be far apart in the trace norm and the verifier accepts.  If the states are close together then the prover will be correct about half of the time and the verifier rejects.\footnote{
A more sophisticated argument involves first amplifying to obtain a pair of circuits with $\alpha$ close to $0$ and $\beta$ close to 1 and then playing the game once.  Then completeness is close 1, soundness is close to $\frac{1}{2}$ and only two messages are required.  See Watrous~\cite{Watrous:2002:Limits-on-the-p} for details.
}

Now we consider two unitaries $U_{0}$ and $U_{1}$ which differ only in their global phase.  In particular, $U_{0} \ket{0 \dots 0} = i U_{1} \ket{0 \dots 0}$.  Clearly the two states must be completely indistinguishable since $\rho_{0} = \rho_{1}$.  However, if we apply the real simulation and pass the ``extra'' qubit to the prover, the states are orthogonal and the prover can distinguish them perfectly!  Indeed, according to lemma~\ref{lemma:propertiesofR}
\begin{eqnarray*}
R(\bra{0\dots0} U^{\dagger}_{0}) R(U_{1} \ket{0\dots 0}) & = & 
\bra{0 \dots 0} U^{\dagger}_{0} U_{1} \ket{0 \dots 0} \\
& = & \text{Re}\, i \\
& = & 0.
\end{eqnarray*}

A closer inspection reveals that $R(U_{0} \ket{0\dots 0}) = (X \otimes I) R(U_{1} \ket{0 \dots 0})$.  Thus the prover needs only to measure the ``extra'' qubit in the $Z$ basis  to distinguish the states.  The reason that the previous proof fails for this construction is that the prover's measurement on the ``extra'' qubit does not commute with measurement in the $V^{\dagger}ZV$ basis, so we can no longer view the verifier's actions as a mixture of the original and complex conjugate protocols  (they share the ``extra'' qubit).  In the construction with two ``extra'' qubits the verifier keeps one qubit, so \emph{all} operations, both the verifier's and the prover's, commute with measurement in the $V^{\dagger}ZV$ basis.

\end{document}